\documentclass[journal]{IEEEtran}
\usepackage{amssymb}
\usepackage{graphicx}
\usepackage{url}
\usepackage{amsmath}
\usepackage{subfig}
\usepackage{comment}

\newcommand{\ud}{\mathrm{d}}

\ifCLASSINFOpdf

\else

\fi

\begin{document}

\title{Correction Algorithm of Sampling Effect and Its Application}

\author{Yunqi~Sun,
        Jianfeng~Zhou

\IEEEcompsocitemizethanks{\IEEEcompsocthanksitem Jianfeng Zhou is with the Department of Engineering Physics, Tsinghua University, Beijing, 100084, China and Xingfan Information Technology Co., Ltd.(Ningbo), Zhejiang 315500, China. E-mail: jianfeng\_zhou@qq.com. 

\IEEEcompsocthanksitem Yunqi Sun is with the Department of Physics, Tsinghua University, Beijing, 100084, China. E-mail: yq-sun17@mails.tsinghua.edu.cn.}

\thanks{Manuscript received January 21, 2021; revised August 26, 2021. This work was supported in part by the National Program on Key Research and Development Project (Grant No. 2016YFA0400802), and in part by the National Natural Science Foundation of China (NSFC) under Grant No. 11173038 and No. 11373025.
}}

\markboth{IEEE TRANSACTIONS ON IMAGE PROCESSING,~Vol.~14, No.~8, August~2015}%
{Shell \MakeLowercase{\textit{et al.}}: Bare Demo of IEEEtran.cls for IEEE Journals}

\maketitle

\begin{abstract}
The sampling effect of the imaging acquisition device is long considered to be a modulation process of the input signal, introducing additional error into the signal acquisition process. This paper proposes a correction algorithm for the modulation process that solves the sampling effect with high accuracy. We examine the algorithm with perfect continuous Gaussian images and selected digitized images, which indicate an accuracy increase of $10^{6}$ for Gaussian images, $10^2$ at $15$ times of Shannon interpolation for digitized images, and $10^{5}$ at $101$ times of Shannon interpolation for digitized images. The accuracy limit of the Gaussian image comes from the truncation error, while the accuracy limit of the digitized images comes from their finite resolution, which can be improved by increasing the time of Shannon interpolation.
\end{abstract}


\IEEEpeerreviewmaketitle

\section{Introduction}
\IEEEPARstart{T}{he} sampling effect describes the average process that the image sensor works as an integral sampler which makes the recorded signal deviate from the input signal. This effect is referred to in some studies as part of the total modulation transfer function(MTF) of an image acquisition process\cite{2004Modeling}\cite{2001Modulation}\cite{Feltz:90}\cite{4758218}. In high accuracy image acquisition, researchers proposed compensation algorithms for the total MTF to improve the imaging quality\cite{Li:17}\cite{Choi:08}\cite{10.1117/12.56012}. However, the study for the feature of the integral sampler's sampling effect is long considered to be an inevitable mistake.

Substituting the impulse signal with the integral signal can cause over $1\%$ relative errors of the recorded signal. For most imaging systems, the difference between the integral sample and the impulse sample are negligible, and thus the error is tolerable. However, for imaging systems demanding high accuracy signal recording, this substitution causes severe accuracy loss in the recording process. In the subject of Fourier profilometry, the sampling effect causes errors in the depth measurement\cite{SU2001263}\cite{ZUO201870}\cite{Yin:19}. The demand for high accuracy imaging analysis urges a recovery algorithm for the integral sample signal to understand the original signal better.

This paper studied the integral sampling effect from a basic image sensor model and gave the direct connection between the impulse sample and the integral sample by the Fourier analysis method. We proved that the sampling effect causes accuracy loss in signal amplitude. For images with appropriate sampling frequency and spatial feature, we put forward a matrix equation for calculating the accurate impulse sample from the integral sample, which decreases the error between the recorded signal and the impulse sample by $10^{12}$ at most. By testing the matrix equation on both simulated and actual images, We further improve the sensor model and combine the matrix equation with the current imaging scheme.

\section{Models for image sensors and sampling process}
The sampling theory and the mathematics for the sampling process are discussed profoundly\cite{2004Modeling}\cite{2001Modulation}\cite{Feltz:90}\cite{modelimgsys}\cite{1984Modulation}. Based on this theory, we now consider an ideal image sensor array. The single image sensor unit has a size of $2L\times 2L$, and its photon sensitive area has a size of $2(L-d)\times 2(L-d)$, locating at the centre of the sensor unit. The image sensor array is a duplication of the image sensor unit, which forms the basic geometric model for the ideal image sensor array.

\subsection{Models for Integral Sampling Process}
We introduce a model for describing the photon flux distribution on an imaginary plane that coincides with the image sensors. We define the optical centre of the imaging system as the centre of the coordinate system $(u,v)$ on this imaginary plane. Since the recording process accumulates all the incident photons on a single image sensor unit, the readout signal should be proportional to the integral of incident flux within shutter time, which gives an ideal integral sampling process:
\begin{equation}
\label{eqn_isample}
	\tilde{I}_{mn} = \iint I(u,v)S(m-u,n-v)\ud u\ud v,
\end{equation}
where $\tilde{I}_{mn}$ is the integral flux on the sensor unit on the m$th$ row and n$th$ column, $I(u,v)$ is the photon flux distribution before they reach the sensor, $S(u,v)$ is the shape function of the single image sensor unit with $S(u,v) = 1$ for $(u,v)$ in the boundary of the unit and $S(u,v) = 0$ for other situation. We assume the incident flux is time-independent and dismiss the shutter time influence on the sampling for convenience.

We further abstract an integral model before the signal sampled by the sensor unit. Recall that a typical impulse sample process is $X_{mn} = X(u=mT,v=nT)$ with $T$ is the sampling interval for a given signal $X(u,v)$. We can consider the $\tilde{I}_{mn}$ as the impulse sample of a certain signal $\tilde{I}(u,v)$. The definition of $\tilde{I}(u,v)$ is given directly as,
\begin{equation}
\label{eqn_isample}
	\tilde{I}(u,v) = \iint I(s,t)S(u-s,v-t)\ud s\ud t = I(u,v)\ast S(u,v).
\end{equation}

The Shannon Sampling Theorem gives the relation between the impulse signal and its corresponding continuous signal. This can be easily applied to $I(u,v)$ and $I_{mn}$ as,
\begin{equation}
\label{eqn_sampp}
    I(u,v) = \sum_{mn}I_{mn}{\rm sinc}(\frac{u-mD}{D}){\rm sinc}(\frac{v-nD}{D})
\end{equation}
where $D$ is the distance between 2 neighbouring sample points. ${\rm sinc}(x) = \sin(\pi x)/(\pi x)$. $I(u,v)$ is the flux distribution and $I_{mn}$ is the impulse sample of $I(u,v)$. In the geometric model for sensor unit, $D=2L$ is the length of a single sensor unit.

\begin{figure}[]
	\centering
	\includegraphics[scale = 0.7]{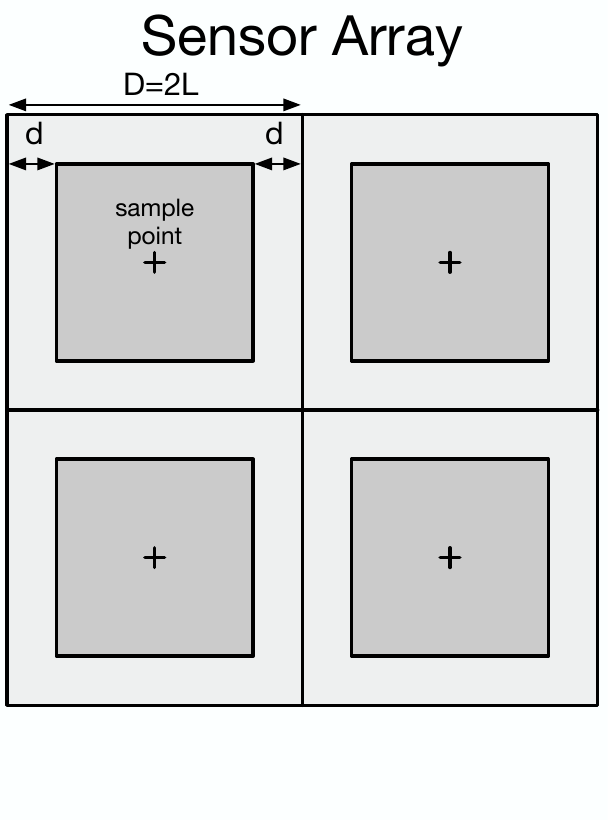}
	\caption{Geometric model for the sensor unit array. The grey area is the photon sensitive area of the sensor unit.The cross points are the centers of the sensor unit.}
	\label{fig_unit}
\end{figure}

Combine \ref{eqn_isample} and \ref{eqn_sampp} gives the direct connection between the integral sample $\tilde{I}_{mn}$ and the flux distribution $I(u,v)$, such that,
\begin{eqnarray}
\label{eqn_connec}
\begin{aligned}
    \tilde{I}(u,v) &= \iint I(s,t)S(u-s,v-t)\ud s\ud t\\
    &= \iint \sum_{mn}I_{mn}{\rm sinc}(\frac{s-mD}{D}){\rm sinc}(\frac{t-nD}{D})\cdot\\
    & S(u-s,v-t)\ud s\ud t\\
    &=\sum_{mn}I_{mn}\iint {\rm sinc}(\frac{s-mD}{D}){\rm sinc}(\frac{t-nD}{D})\cdot\\
    & S(u-s,v-t)\ud s\ud t,
\end{aligned}
\end{eqnarray}
where the summation and the integral symbol are exchangeable because the integral is finite in \ref{eqn_isample} within the size of the sensor array, which is defined as a set $\Lambda$. We define a shape integral $S_{mn}(u,v) = \iint_{\Lambda} {\rm sinc}(\frac{s-mD}{D}){\rm sinc}(\frac{t-nD}{D})S(u-s,v-t)\ud s\ud t$. In this notation, \ref{eqn_connec} can be simplified and we can direct write its discrete version by substituting $(u,v)$ at its sample points $(i,j)$ to get $\tilde{I}_{ij}$, such that,
\begin{equation}
\label{eqn_disc}
    \tilde{I}_{ij} = \sum_{mn} I_{mn} S_{ijmn},
\end{equation}
where 
\begin{eqnarray}
\begin{aligned}
    S_{ijmn} &= S_{mn}(u=iD,v=jD) \\
    &= \iint_{\Lambda} {\rm sinc}(\frac{s-mD}{D}){\rm sinc}(\frac{t-nD}{D})\cdot \\
    & S(iD-s,jD-t)\ud s\ud t.
    \end{aligned}
\end{eqnarray}

This equation indicates the direct relation between the impulse sample signal $I_{mn}$ and the integral sample signal $\tilde{I}_{mn}$. The integral sample is different from the impulse sample by a shape matrix $S_{ijmn}$, which is determined by the geometric parameters of the image sensor unit.

\subsection{Practical Method for Integral Sampling Correction}
We provide a practical method for solving the sampling process with an image sensor unit with a simple geometric structure. Consider a rectangle shape function with given gap around it as is shown in \ref{fig_unit}, the shape function is $Rect(x,y) = 1$, for $|x|<1/2$, and $|y|<1/2$, where $x = u/(L-d)$, $y = v/(L-d)$. The shape matrix of this rectangle is,
\begin{equation}
    \label{eqn_rectshape}
    \begin{aligned}
    S^R_{ijmn} &= \iint {\rm sinc}(\frac{s-mD}{D}){\rm sinc}(\frac{t-nD}{D})\cdot \\
    & {\rm Rect}(\frac{i-s}{L-d},\frac{j-t}{L-d})\ud s\ud t,
    \end{aligned}
\end{equation}
which can be fully determined with the given geometric parameters of the sensor unit. We should notice that $S^R_{ijmn}$ can be divided into the product of two identical symmetric matrix $\mathbb{R}$ with $R_{mi} = \int {\rm sinc}(\frac{s-mD}{D}){\rm Rect}(\frac{i-s}{L-d})\ud s$, and $S^R_{ijmn} = R_{mi}R_{nj}$. \ref{eqn_disc} is then simply $\tilde{I}_{ij} = \sum_{mn}I_{mn}R_{mi}R_{nj}$, or $\mathbb{\tilde{I}} = \mathbb{R}\cdot\mathbb{I}\cdot\mathbb{R}$.
The reconstruction method can be noted as,
\begin{equation}
    \label{eqn_recmethod}
    \mathbb{I} = \mathbb{R}^{-1}\cdot\mathbb{\tilde{I}}\cdot\mathbb{R}^{-1},
\end{equation}
where the inverse matrix of $\mathbb{R}$ should exist to assure the correction method.

\subsection{Spatial Frequency Constraints}
\label{subsec_cons}
The application of Shannon Sampling Theorem into the correction method for the sampling effect gives an inherent constrain on the spatial frequency of $\tilde{I}_{mn}$. The theorem only treats a band-limited signal with its highest spatial frequency $f_c$ smaller than half of the sampling rate of the system $f_N$\cite{Boreman1999Oversampling}. The Nyquist sampling rate $f_N$ of the imaging system is determined by the imaging device and the image sensor, which is $f_N = 1/D$. For a imaging device with a Gaussian PSF with $p(u,v) = 1/(2\pi\sigma^2)\exp(-(u^2+v^2)/\sigma^2))$, the spectrum of the PSF is its Fourier Transform $P(U,V)$, which is still a Gaussian distribution with $P(U,V) = \exp(-2\pi^2\sigma^2(U^2+V^2))$. For $|U|>5/(2\pi\sigma)$ and $|V|>5/(2\pi\sigma)$, we can consider $P(U,V)\sim 0$. This gives the bandwidth of the Gaussian PSF with $f_c = 5/(2\pi\sigma)$. The Nyquist Sampling Theorem requires,
\begin{equation}
    f_c<\frac{f_N}{2},
\end{equation}
which is equivalent to
\begin{equation}
    \frac{\sigma}{D}>\frac{5}{\pi},
    \label{eqn_cons}
\end{equation}
note that $\sigma/D$ is the equivalent standard deviation for an dimensionless imaging system.

To design an imaging system that meets the Nyquist Sampling Theorem, \ref{eqn_cons} is an essential constraint.

\section{Simulations and Applications}
To verify the influence of the integral sample and the correction algorithm, we generate a series of simulations and test the correction method on them. The simulation procedure includes the following steps. First, we set a shift-invariant imaging system with a given point spread function. Then we generate the 2-D object, which is a 2-D flux distribution on the object plane. After the imaging system images the 2-D object, we sample the signal by impulse sample method and integral sample method and apply the correction algorithm to the integral sample signal to acquire a signal recovered of the sampling process.

\subsection{Gaussian Point Spread Function Simulation}
We simulate an imaging system which has an Gaussian point spread function(PSF). The sensor unit has a dimensionless size of $1\times 1$(which means $D=1$), and the PSF of the system is $p(u,v) = \exp(-(u^2+v^2)/(2\sigma^2))$ with $\sigma = 3$. The object plane has two ideal point sources with a unified flux of 1. We simulate the imaging process of this system with the given object plane and sample the images by the integral sampler and impulse sampler separately. 

Fig. \ref{fig_gaussian} shows the sampled image and the recovered image from the integral sampler. This simulation indicates that the sampling effect from substituting the impulse sampler by an integral sampler can cause significant relative error up to $1\%$ at the centre of the incident flux. By applying the recovery algorithm on the integral sample image, we recover an accurate flux distribution of the two-point sources after the imaging device and reduce the relative error to $10^{-8}$, which is $10^6$ better than the integral sample image. 

Also, the error image itself shows a fringe pattern at its corner, which is spatially connected to the point source at $(26,26)$ in the image's coordinate system. This pattern is a Gibbs phenomenon caused by the discontinuous components from the point source at $(26,26)$, and it contributes the most in the error component in the recovery method.

\subsection{Real Image Simulation}
\begin{figure}[]
	\centering
	\subfloat[]{
		\includegraphics[width=0.35\linewidth]{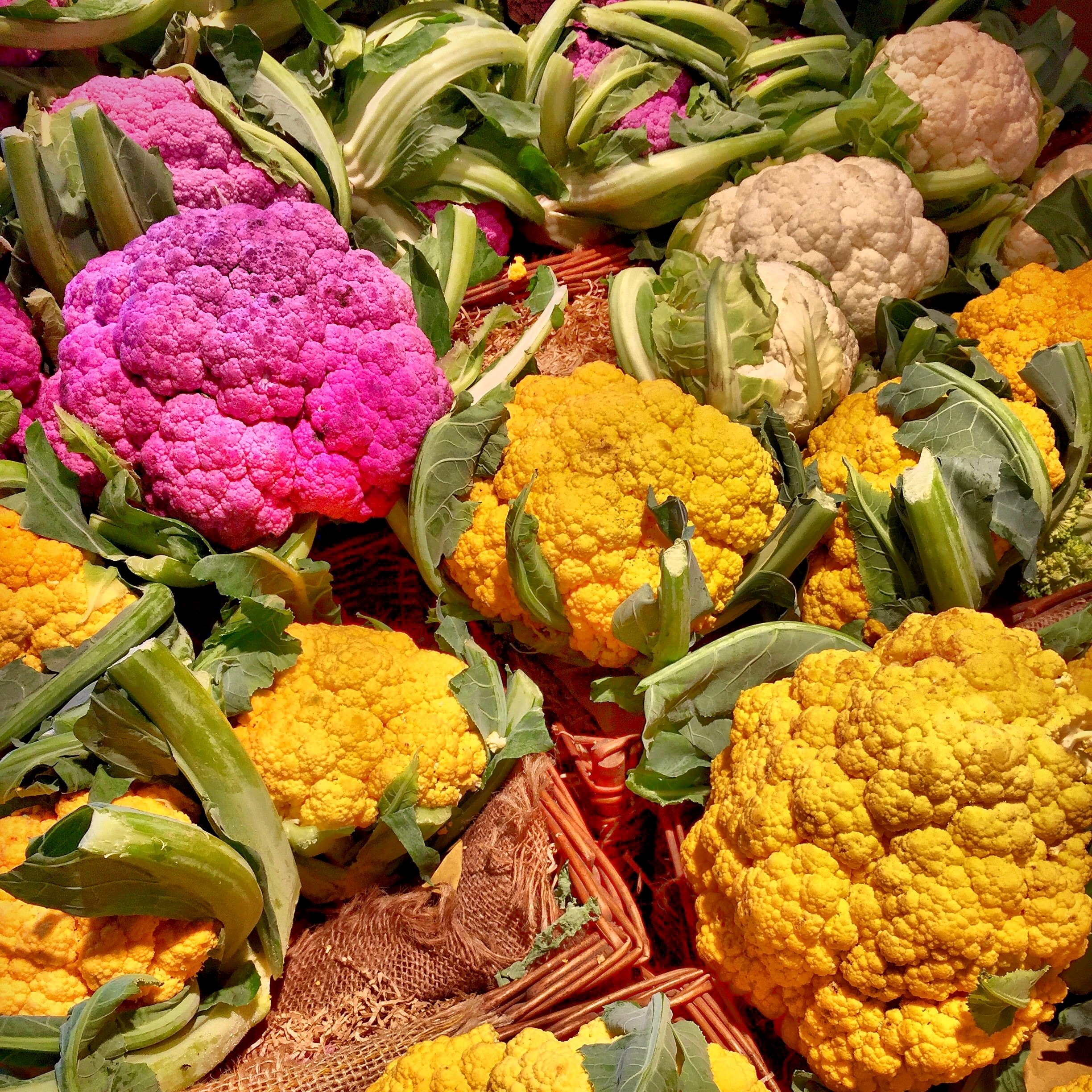}}
		\quad
	\subfloat[]{
		\includegraphics[width=0.35\linewidth]{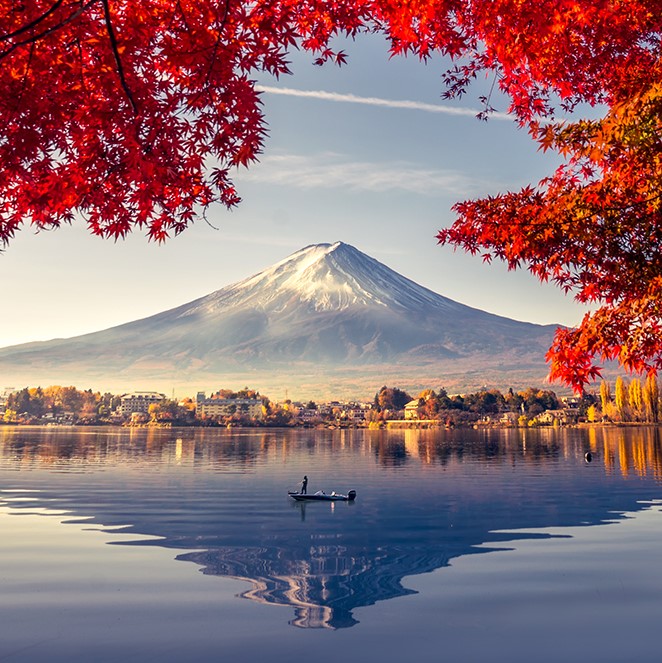}}
	\caption{Two random images from the internet. (a) is a basket of vegetables. (b) is Mount Fuji.}
	\label{fig_realimg}
\end{figure}

We further illustrate the algorithm by two real images (Fig. \ref{fig_realimg}). These two images represent the flux distribution of the 2-D object plane and form two ideal images after the imaging system. The imaging system has a Gaussian PSF ($\sigma = 2$) which meets the Nyquist sampling theorem. 

\begin{figure*}[]
	\centering
	\includegraphics[scale = 0.6]{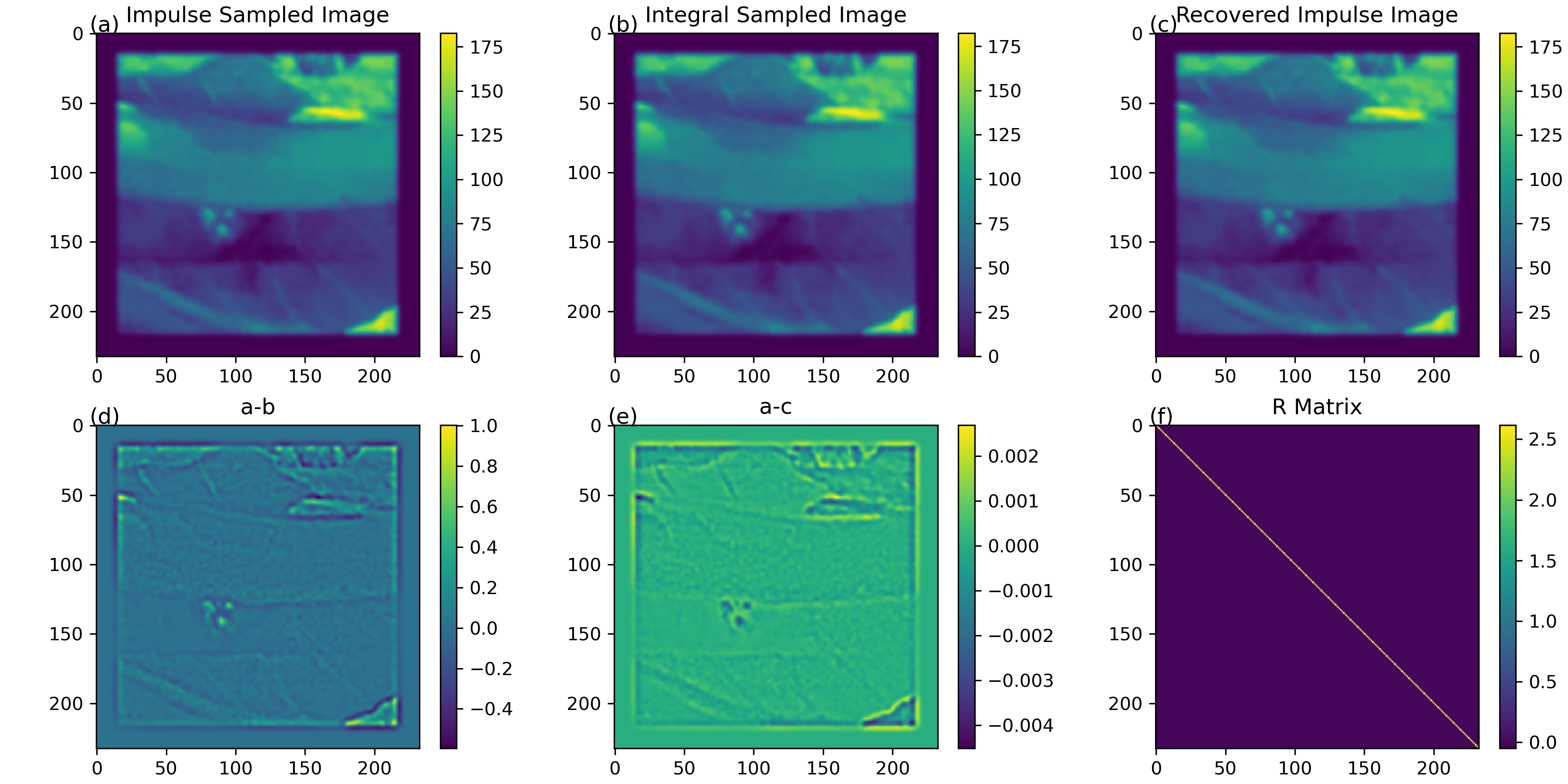}
	\caption{Simulated image vegetable and its two samples (a) and (b), together with the recovered image (c) from the integral sample shown in the pixel coordinate. (a) is a $201\times201$ area from Fig. \ref{fig_realimg} (a). (d) shows the error image of two images by different sample methods in (a) and (c). (e) shows the error image of the impulse sampled image (a) and the recovered image (c). (f) is the $\mathbb{R}$ matrix for this sampling system.}
	\label{fig_realrec1}
\end{figure*}

\begin{figure*}[]
	\centering
	\includegraphics[scale = 0.55]{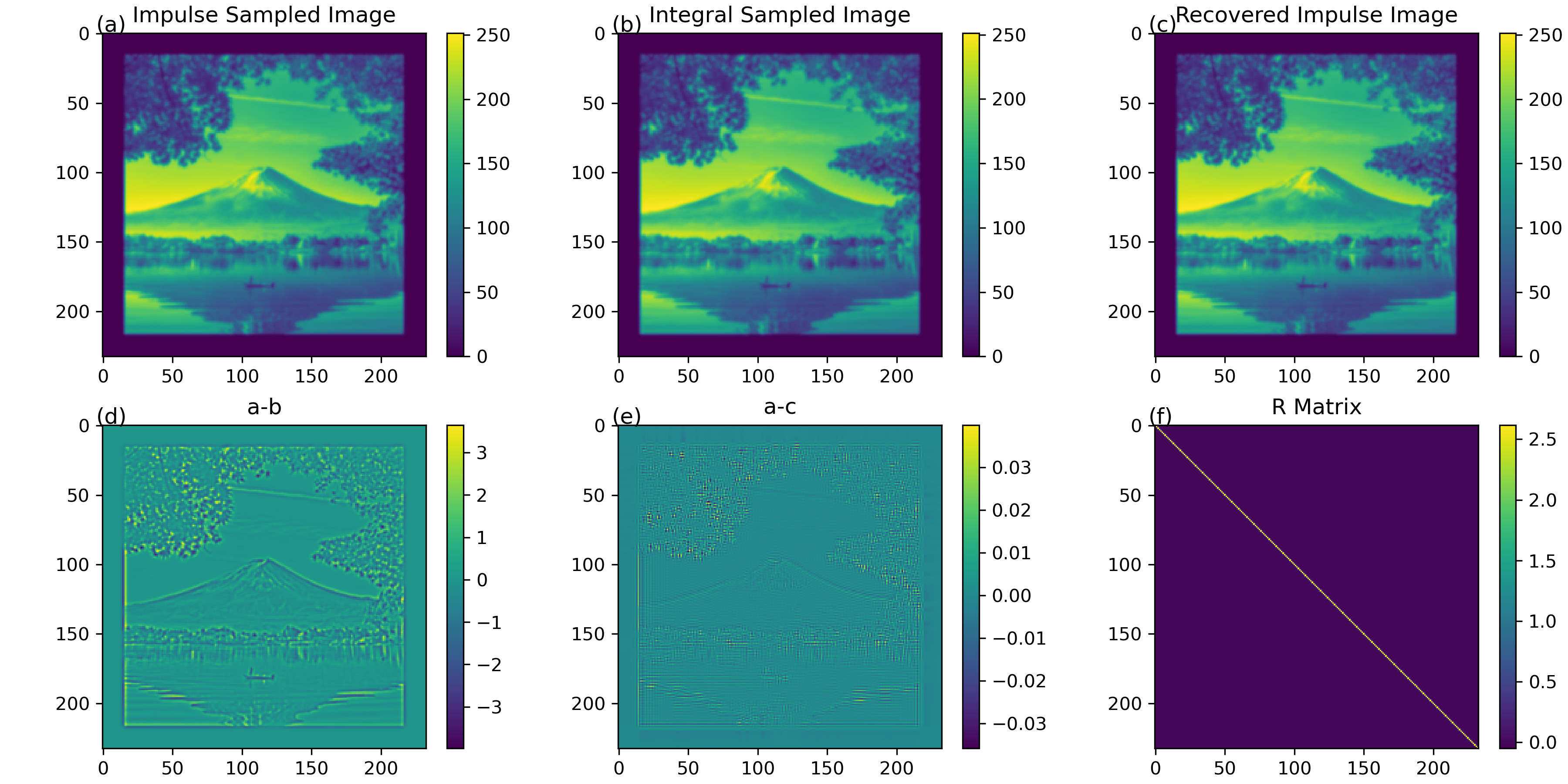}
	\caption{Simulated image Mount Fuji and its two samples (a) and (b), together with the recovered image (c) from the integral sample shown in the pixel coordinate. (a) is the simulated image of 2-D object Fig. \ref{fig_realimg} (a). (d) shows the error image of two images by different sample methods in (a) and (c). (e) shows the error image of the impulse sampled image (a) and the recovered image (c). (f) is the $\mathbb{R}$ matrix for this sampling system.}
	\label{fig_realrec2}
\end{figure*}

 After the 2-D objects go through the imaging device, we apply $15$ times of Shannon's interpolation\cite{1697831} to increase their resolution to improve the accuracy of a simulated integral sampling process. We sample the two ideal images using the impulse and integral sample methods and apply the recover algorithm on both integral sampled images. The results are in Fig. \ref{fig_realrec1} and Fig. \ref{fig_realrec2}.

\begin{table*}[]
\centering
\begin{tabular}{ccccc}
\hline
                            & Vegetable Integral & Vegetable Recovered & Mount Fuji Integral & Mount Fuji Recovered  \\ \hline
Maximum Pixel Deviation(\%) & $5$                & $2.7\times10^{-4}$  & $1.4$               & $2.3\times10^{-3}$ \\ \hline
Total Energy Deviation(\%)  & $0.02$             & $5.3\times10^{-9}$  & $0.01$              & $1.8\times10^{-7}$ \\ \hline
\end{tabular}
\caption{Deviation from the Impulse Sample Images with Integral Sample Images and Recovered Images.}
\label{table_comp}
\end{table*}

\begin{figure*}[]
	\centering
	\includegraphics[scale = 0.55]{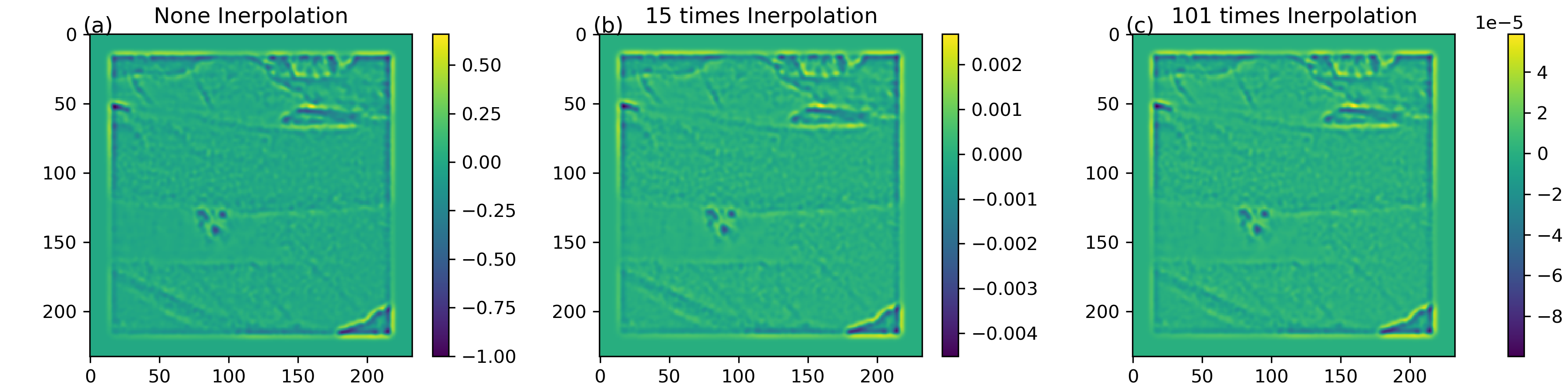}
	\caption{Recovered error images by none, $1$ time and $101$ times Shannon interpolation. (a) is none interpolation. (b) is $15$ times interpolation. (c) is $101$ times interpolation. The accuracy improved $250$ times as the interpolation increased from $1$ time to $15$ times, and it only improved $50$ times as the interpolations increased from $15$ times to $101$ times.}
	\label{fig_errcmp}
\end{figure*}

The results in Fig. \ref{fig_realrec1} and Fig. \ref{fig_realrec2} show the process that the 2D objects are sampled and recovered from the integral average. The error images illustrate that the recovery algorithm greatly improves the readout image's accuracy to about at least 100 times better than the integral sample image. Detailed research shows that the improvement is strongly related to the integral's numerical accuracy. As we apply Shannon interpolation with a higher time, the recovery becomes more accurate to the impulse sampled image, which is shown in Fig. \ref{fig_errcmp}. More details for the comparison in Fig. \ref{fig_realrec1} and Fig. \ref{fig_realrec2} are presented in table \ref{table_comp}. We define the maximum pixel deviation and the total energy deviation for further analysing the results. The definition of the maximum pixel deviation(MPD) is
\begin{equation}
    {\rm MPD} = \frac{{\rm max}(I_{k}-I_{\rm PS})}{{\rm max}(I_{\rm PS})},
\end{equation}
where $I_{PS}$ refers to the impulse sampled image, and $I_{k}$ refers to the integral sampled image and the recovered impulse sampled image respectively. The definition of the total energy deviation(TED) is
\begin{equation}
    {\rm TED} = \frac{\sum(I_{k}-I_{\rm PS})^2}{\sum I^2_{\rm PS}}.
\end{equation}

Results in Table \ref{table_comp} indicates that the integral sampled image has about $0.01\%$ total energy error to the ideal impulse sampled image while the recovery method reduces the total energy deviation to only $10^{-9}$ at $15$ times Shannon interpolation. Also, the maximum pixel deviation is reduced from $5\%$ to $10^{-6}$ at $15$ times Shannon interpolation. This algorithm is a major improvement for image measurement for it enables the detection of non-bias photon energy and accurate impulse sampling for normal image devices with an integral sampler.

\section{General Deconvolution Solver for Band-Limited Signal}
Recall equation \ref{eqn_isample}, we find that the deduction of the reconstruction method for the sampling effect is essentially an reconstruction method for the convolution equation \ref{eqn_isample}.

\section{Discussion}
\begin{figure}[]
	\centering
	\includegraphics[scale = 0.6]{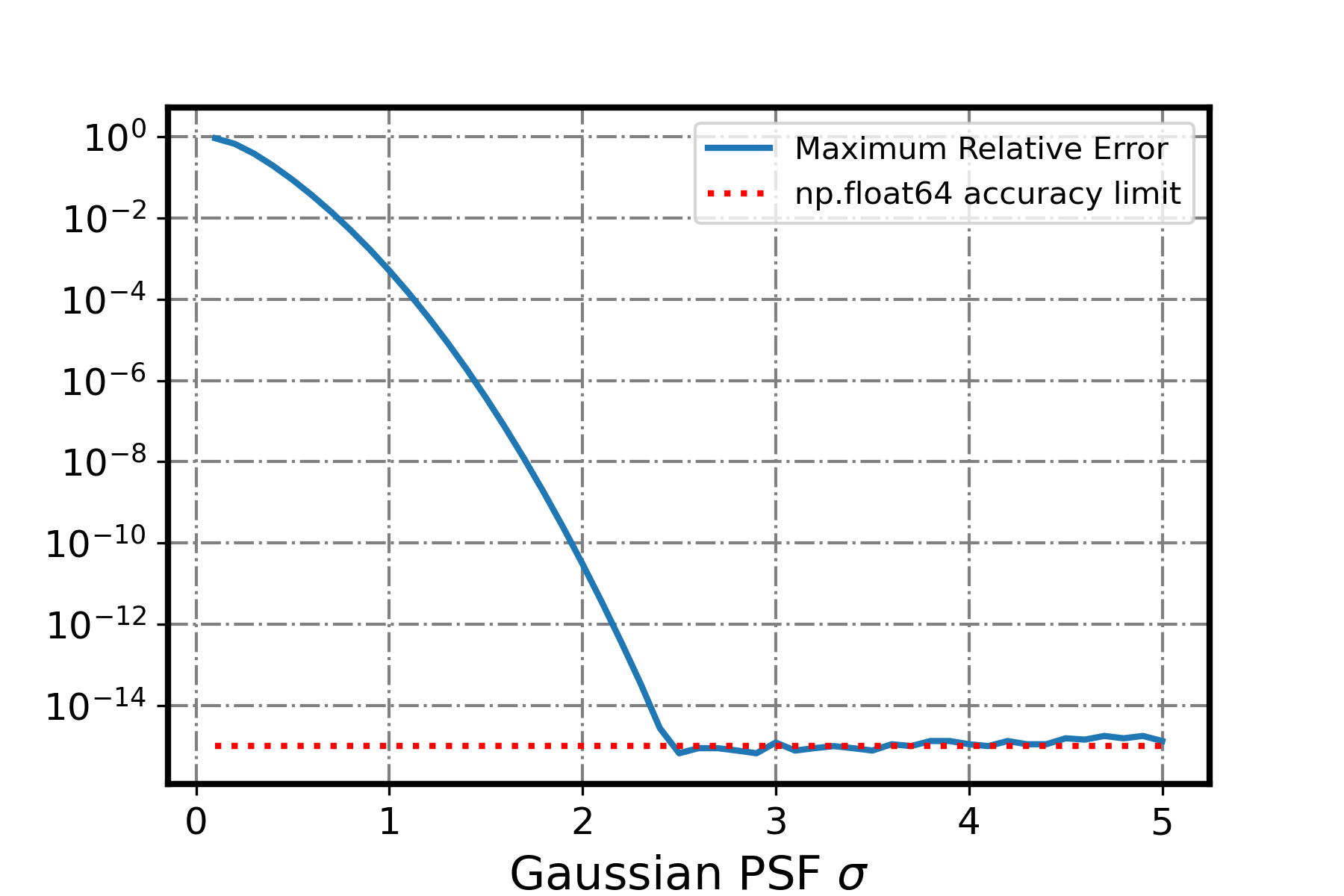}
	\caption{Error analysis about the PSF size and the sampling rate. The maximum relative error decreases rapidly to the numerical accuracy limit of numpy's float64, which is about $10^{-15}$.}
	\label{fig_erra}
\end{figure}

Two key factors determine the improvement of an integral sample signal. One is the spatial frequency response of the imaging system. A PSF that fully meets the Nyquist Sampling Theorem would provide the most accurate result when calculating the shape matrix $\mathbb{R}$. The design of high accuracy imaging system should consider the connection between the imaging device and the image sensor as is discussed in \ref{subsec_cons}. Fig. \ref{fig_erra} shows the relation between the relative error and the system's $\sigma$.

Fig. \ref{fig_erra} also indicates that the demand for high-accuracy imaging systems should focus on producing image sensors with smaller sizes instead of generating imaging devices with smaller PSF. As \ref{subsec_cons} discussed, an appropriate $\sigma$ which is about $2$ times of the pixel's length is more appropriate for designing an high-accuracy image acquisition system than a small $\sigma$.

\begin{figure*}[]
	\centering
	\includegraphics[scale = 0.55]{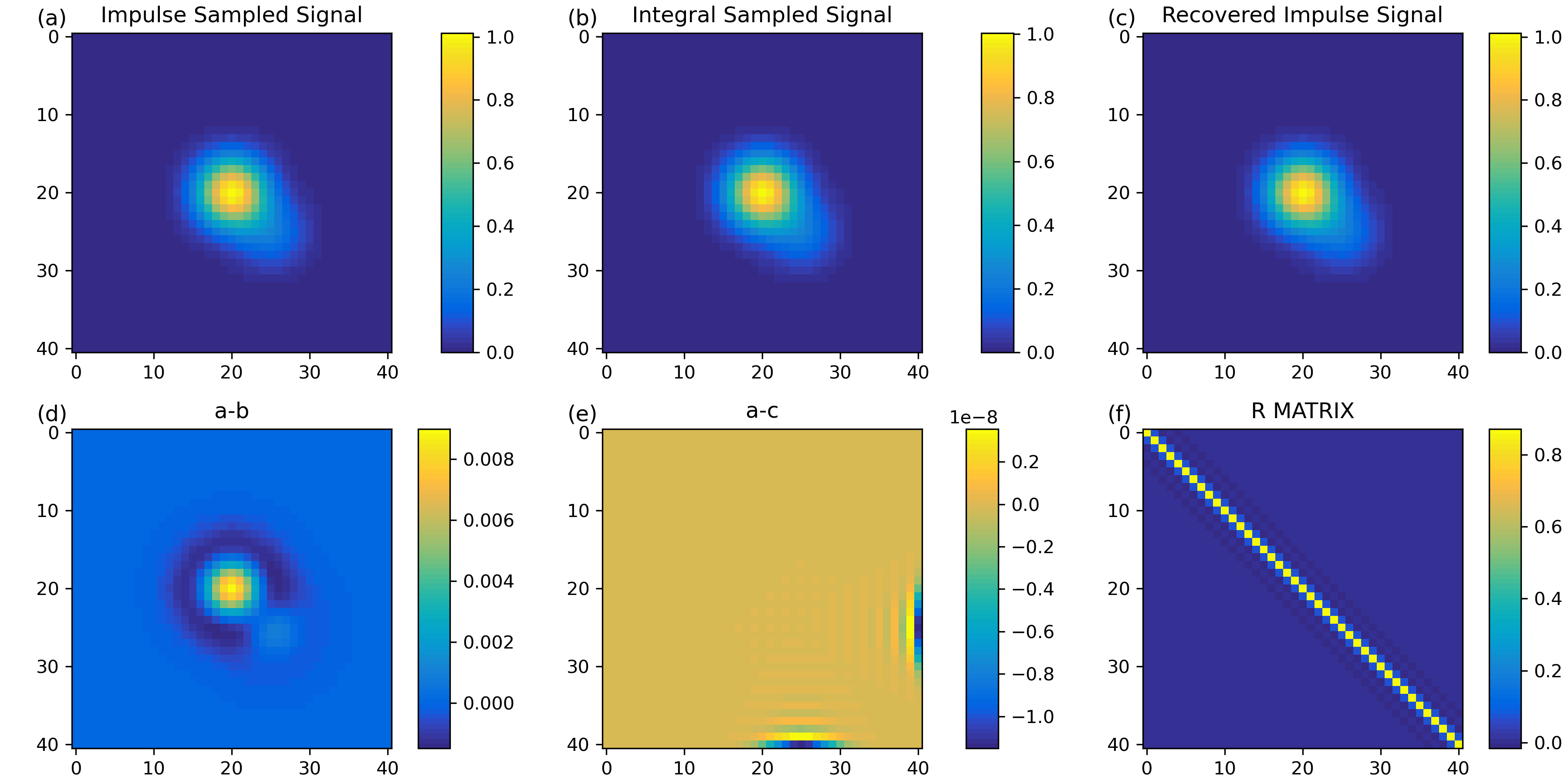}
	\caption{Simulations for two-point sources collected by an imaging system with a Gaussian PSF shown in the pixel coordinate. The point sources are at $(21,21)$ and $(26,26)$ in pixel coordinate on the object plane. (a) is the impulse sample signal. (b) is the integral sample signal. (c) shows the recovered impulse signal from (b). (d) is the error image of (a) and (b). (e) is the error image of (a) and (d). (f) is the $\mathbb{R}$ matrix for this system.}
	\label{fig_gaussian}
\end{figure*}

The other is the simple fact that the discrete Fourier transform in the algorithm requires the signals to be periodic and continuous at their border. The discontinuous components contribute the main error in our recovery algorithm, which can be directly spotted from the Gibbs phenomenon in Fig. \ref{fig_gaussian}. To reduce the error induced by the discontinuous components, it is feasible to apply additional Gaussian low-pass filter on the integral sampled signal. 

\section{Conclusion}
This paper discussed the mathematical model for the image sensors as an integral sampler and the relative error introduced by the integral sampling method. The relative error caused by the integral sampler relates strongly to the size of the sensor's unit compared to the PSF of the imaging device. We provide a thorough correction algorithm for imaging systems with proper PSF for reducing the relative error regardless of the sensor's size. By testing the algorithm on actual images in simulated imaging system, we validate the algorithm's accuracy, which is partially influenced by the accuracy of the simulated integral sampler and the Shannon's interpolation. We also discussed the essential demand of acquiring high-accuracy imaging systems. Instead of decreasing the PSF's size, decreasing the imaging sensor's size is more critical in improving the system's accuracy. In general, this algorithm has the potential of broaden the accuracy of energy and flux detection for normal imaging systems with integral sampler, and is also a possible method for improving accuracy for general integral sampler.

\section*{Acknowledgment}

\ifCLASSOPTIONcaptionsoff
  \newpage
\fi

\bibliographystyle{ieeetr}
\bibliography{ref}
\clearpage
\begin{IEEEbiography}[{\includegraphics[width=1in,height=1.25in,clip,keepaspectratio]{figure/syq.pdf}}]{Yunqi Sun}
Yunqi Sun received a Bachelor's degree in Engineering Physics from Tsinghua University in 2017. He is currently pursuing his Ph. D. degree in Astrophysics at Tsinghua University.
His research interests include image distortion correction, image super-resolution, and image reconstruction algorithms.
\end{IEEEbiography}

\begin{IEEEbiography}[{\includegraphics[width=1in,height=1.25in,clip,keepaspectratio]{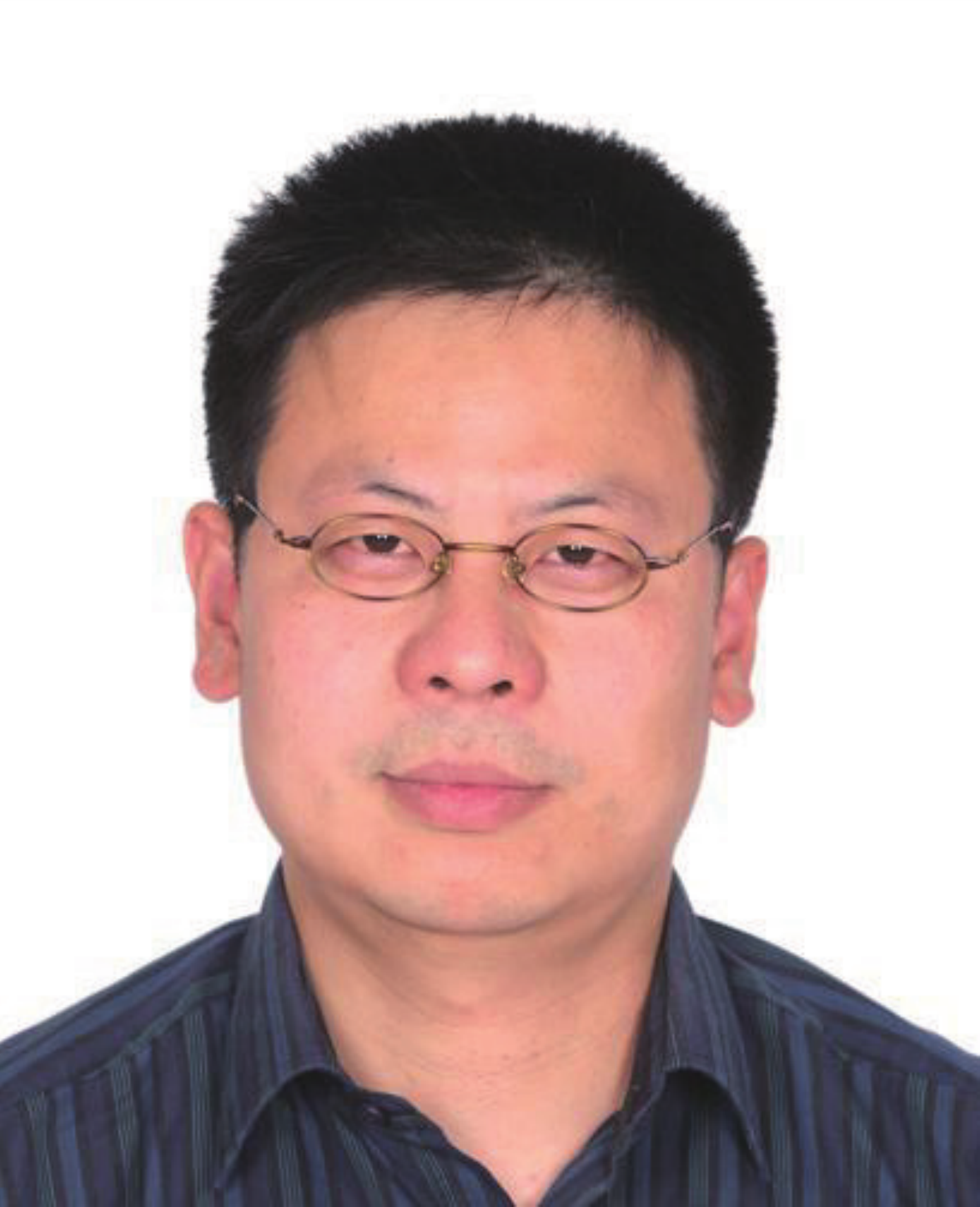}}]{Jianfeng Zhou}
Jianfeng Zhou received the B.Sc. degree in Geophysics from the University of Science and Technology of China in 1995, and the M.Sc. and Ph. D. degrees in Astrophysics from Shanghai Astronomical Observatory, Chinese Academy of Sciences, in 1998 and 2001, respectively. From 2001 to 2004, he was a post-doctoral researcher at the Center for Astrophysics at Tsinghua University in China and then joined the faculty there. He is currently an Associate Professor.

His technical interests include imaging and image reconstruction methods and astrophysics.
\end{IEEEbiography}
\ifCLASSOPTIONcaptionsoff
  \newpage
\fi

\end{document}